\documentclass{IEEEtran}
\usepackage{amsmath,amsfonts}
\usepackage{array}
\usepackage{textcomp}
\usepackage{stfloats}
\usepackage{url}
\usepackage{verbatim}
\usepackage{graphicx}
\usepackage{bm}
\usepackage{booktabs}
\usepackage{cite}
\usepackage{hyperref}
\usepackage{array}
\usepackage{multirow}
\usepackage{multicol}
\usepackage{arydshln}
\usepackage{booktabs}  % for table
\usepackage{makecell}
\usepackage{siunitx}
\usepackage{xcolor}
\usepackage{pgfplots}
\usepackage{tikz}
\usepackage[end]{algpseudocode}  % for algorithm
\usepackage{algorithmicx}
\usepackage{algorithm}
\usepackage{tikz}
\usetikzlibrary{arrows.meta, positioning, shapes.geometric}

\def\BibTeX{{\rm B\kern-.05em{\sc i\kern-.025em b}\kern-.08em
    T\kern-.1667em\lower.7ex\hbox{E}\kern-.125emX}}
\usepackage{balance}
\usepackage{xcolor}

\newif\ifrev \revfalse
\newcommand{\rev}[1]{\ifrev\textcolor{red}{#1}\else#1\fi}

\begin{document}
\title{Adaptive Polynomial Chaos Expansion for Uncertainty Quantification of Radio Wave Propagation over Irregular Terrains}
% \title{Adaptive Polynomial Chaos Expansion for Uncertainty Quantification and Sensitivity Analysis of Parabolic Equation Modeling of Radio Wave Propagation over Irregular Terrain}
% \title{A Novel Adaptive Polynomial Chaos Expansion for Uncertainty Quantification of Parabolic Wave Equation Modeling of Radio Wave Propagation over Irregular Terrain}
% \title{Uncertainty Quantification of Radio Wave Propagation over Irregular Terrains Using Adaptive Polynomial Chaos Expansion}
% \title{Impact of Antenna-Parameter Uncertainty on Path Loss over Irregular Terrain}
\author{Sicheng~An,~\IEEEmembership{Graduate Student Member,~IEEE},
~Lorenzo~Codecasa,~\IEEEmembership{Member,~IEEE},
~Hao~Qin,~\IEEEmembership{Member,~IEEE},
~Xingqi~Zhang,~\IEEEmembership{Senior Member,~IEEE}, and
~Luca~Di~Rienzo,~\IEEEmembership{Senior Member,~IEEE},
\thanks{Manuscript received XX XX, XXXX; revised XX XX, XXXX; accepted XX XX, XXXX. (Corresponding author: Hao Qin)}
\thanks{Sicheng An, Lorenzo Codecasa, and Luca Di Rienzo are with the Dipartimento di Elettronica, Informazione e Bioingegneria, Politecnico di Milano, Milan 20133, Italy (e-mail: sicheng.an@polimi.it; lorenzo.codecasa@polimi.it; luca.dirienzo@polimi.it.)}
\thanks{Hao Qin is with the School of Electronics and Information Engineering, Sichuan University, Chengdu 610065, China (e-mail: hao.qin@scu.edu.cn).}
\thanks{Xingqi Zhang is with the Department of Electrical and Computer Engineering, University of Alberta, Edmonton, AB T6G 2H5, Canada (e-mail: xingqi.zhang@ualberta.ca).}% <-this % stops a space
\thanks{Color versions of one or more of the figures in this paper are available online at http://ieeexplore.ieee.org.}
\thanks{Digital Object Identifier: XXXX}}

\markboth{IEEE Transactions on Antennas and  Propagation}%
{An \MakeLowercase{\textit{et al.}}:}

\maketitle 

\begin{abstract}
Accurate modeling of radio wave propagation over irregular terrains is crucial for designing reliable wireless communication systems in such environments, yet uncertainties in the antenna configuration are not quantified within deterministic models. \rev{This paper develops a fixed number of samples adaptive polynomial chaos expansion (APCE) method for uncertainty quantification (UQ) of radio wave propagation over realistic irregular terrains, where the model evaluations are obtained using a two-way parabolic wave equation (PWE) method. The proposed APCE method is designed to construct a compact and stable PCE model from a prescribed and limited set of simulations. The polynomial basis is enriched using an anisotropic basis extension algorithm driven by variance contributions, while validation behavior and the available sample size are used to control basis growth.} The convergence analysis shows decreasing validation errors and improved robustness as the sampling budget increases, with lower trial-to-trial variability than the baseline adaptive PCE method, which uses the same variance-driven basis extension strategy. For two realistic terrain profiles, the proposed method accurately predicts the mean and the 5-95 percentile range of the path loss, and improves the estimation of the standard deviation compared with standard and sparse PCE, using only 30 PWE simulations. APCE outperforms standard and sparse PCE, with the largest gains observed for the 5th and 95th percentile estimates, and its construction time is close to that of standard PCE and much lower than that of sparse PCE based on least angle regression. As the sample size increases, APCE maintains low errors with reduced trial-to-trial variability.
\end{abstract}


\section{Introduction}
\IEEEPARstart{A}{ccurate} modeling of radio wave propagation over irregular terrains is crucial to developing and optimizing reliable wireless communication systems in such environments \cite{Yagbasan2010Characteristic, Guo2017Pade, Huang2025AGeneralizable}. A variety of numerical techniques, including finite difference time domain (FDTD) methods \cite{Batista2013Ahigh}, integral equation-based methods\cite{Nuallain2005Terrain}, and parabolic wave equation (PWE) methods \cite{Silva2012Terrain}, have been developed to handle these scenarios. Among these approaches, the two-way PWE method has emerged as an accurate and computationally efficient approach for modeling radio wave propagation in such environments \cite{Qin2024Efficient, Guo2020TWo, Qin2023Efficient}. However, accurate numerical simulation requires precise information about antenna configuration parameters, which are often subject to significant uncertainties in practical deployments \cite{IEEEstd1492021, ITU_R_F1336}.

\par To analyze the impact of uncertain parameters in computational electromagnetics problems, various uncertainty quantification (UQ) techniques have been applied, e.g. \cite{Zhu2023Enhanced, An2024Efficient}. The  Monte Carlo (MC) method is widely used due to its straightforward implementation and high reliability, but it typically demands a large number of simulations because of its slow convergence. To address this limitation, polynomial chaos expansion (PCE) methods have been introduced as more efficient alternatives \cite{Maitre2012Spectral, Xiu2010Numerical}. \rev{They have been successfully applied in various UQ problems, including complex-fluid modeling \cite{ElMocayd2021Data}, electromagnetic modeling \cite{Codecasa2016Fast}, and acoustic wave modeling\cite{ElMocayd2020Stochastic}.} More recently, sparse regression techniques have been incorporated into PCE to enhance computational efficiency and robustness to noisy data \cite{Luthen2021sparse, Blatman2011Adaptive}. \rev{
Nevertheless, realistic models do not always lead to strongly sparse PCE coefficients, which can limit the effectiveness of sparse regression techniques. As a consequence, for both classical PCE and sparse PCE, the selection of an appropriate polynomial basis remains a difficult and crucial task \cite{Hampton2018Basis}. Several adaptive PCE (APCE) approaches have therefore been developed to construct problem-dependent polynomial bases \cite{Jakeman2015Enhancing, Hao2025Combined}, and a broader discussion can be found in \cite{Luthen2022Automatic}. However, constructing a compact and stable PCE model from a fixed and limited set of training samples remains challenging.}

% Most recently, sparse regression techniques have been integrated into PCE to enhance computational efficiency and robustness against noise \cite{An2024Efficient, Luthen2021sparse}. In parallel, fully Bayesian formulations of PCE have been introduced, treating the PCE as a Bayesian linear regression model and allowing efficient uncertainty quantification and automatic regularization \cite{Sargsyan2014Dimensionality, Shao2017Bayesian, Burkner2023Afully}.

\par Although recent works have combined PWE methods with different UQ approaches to quantify the uncertainty of radio wave propagation in tunnel environments \cite{An2025Multilevel, Zhang2019Statistical}, challenges remain insufficiently addressed.
% in the context of radio wave propagation over irregular terrains. 
% First, uncertainties in the antenna configuration are often not treated, which may lead to propagation predictions that are overly optimistic and insufficiently robust with respect to realistic installation tolerances.
First, when constructing a surrogate model using PCE, the expansion must be truncated to a finite basis; an inappropriate choice of polynomial terms may lead to poor approximation accuracy or high computational cost. 
Second, the sparsity assumption on the expansion coefficients does not hold universally across all applications.

\par In this paper, we develop a fixed number of samples APCE method for UQ of radio wave propagation over irregular terrains with uncertain antenna configurations under limited training budgets. \rev{The method adopts the anisotropic basis-extension algorithm of \cite{Saturnino2019A} to generate candidate polynomial terms, while a composite heuristic stopping criterion is introduced to control basis growth by combining the sample-to-basis ratio and validation-error convergence. The contribution of this work is therefore the fixed number of samples APCE method for constructing a compact and stable surrogate from a prescribed set of PWE simulations over realistic irregular terrains, rather than the basis extension algorithm itself.} The convergence analysis shows that the leave-one-out validation error generally decreases and becomes more stable as the sampling budget increases. Compared with the APCE method in~\cite{Saturnino2019A}, the proposed stopping criterion leads to lower validation errors and reduced trial-to-trial variability in the considered examples. The proposed APCE is evaluated on two realistic terrain profiles and compared with standard PCE, sparse PCE, and MC reference results. The accuracy is assessed in terms of the mean, standard deviation, and 5th--95th percentile range of the path loss. The results show that the proposed APCE provides accurate and robust statistical estimates under limited sampling budgets.

\par The remainder of the article is organized as follows. Section~\ref{sec: PWE} introduces the deterministic modeling method we used. Section~\ref{sec: adaptive_PCE} presents the proposed APCE approach. Section~\ref{sec: examples} reports numerical results on two realistic terrain profiles, including convergence analysis and uncertainty quantification. Finally, Section~\ref{sec: conclusion} summarizes the main findings and concludes the paper.

% \section{Methodology}
% \label{sec: methods}
\section{Deterministic Modeling}
% \section{Preliminaries}
% \label{sec: preliminaries}
% \subsection{Two-way Parabolic Wave Equation Method}
\label{sec: PWE}

\par
\begin{figure}[htbp]
    \centering
    \includegraphics[width=\columnwidth]{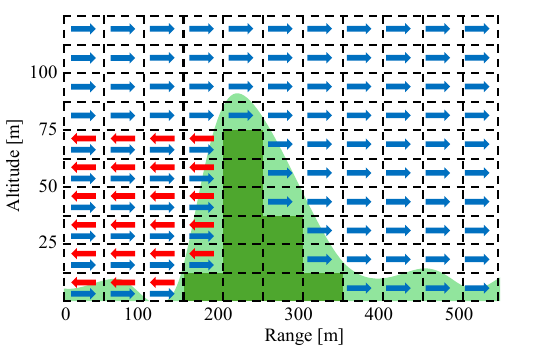}
    \caption{Two-way PWE method over an irregular terrain. The forward (blue) and backward (red) propagating waves are shown; the light-green domain represents the true terrain profile, and the dark-green denotes its discretized approximation.  For clarity, the discretization grid is exaggerated in this illustration; in the actual simulations, a much finer terrain discretization is employed. }
    \label{fig: two_way_PWE}
\end{figure}

 The PWE is derived from the Helmholtz equation \cite{Qin2023Efficient}. For a wave mainly propagating along the $z$-axis, each transverse electric and magnetic field component $\varphi \in \{E_x, E_y, H_x, H_y\}$ satisfies
\begin{equation}
	\left(\frac{\partial^2}{\partial x^2}+\frac{\partial^2}{\partial y^2}+\frac{\partial^2}{\partial z^2}+k_0^2\right)\varphi(x,y,z)=0,
	\label{eq: Helmholtz}
\end{equation}
where $k_0$ is the free-space wavenumber. Expressing $\varphi(x, y, z)$ as
\begin{equation}
	\varphi(x,y,z)=u(x,y,z)e^{-jk_0z},
	\label{eq: solution_Helmholtz}
\end{equation}
where $u(x,y,z)$ is the reduced plane wave solution, and substituting \eqref{eq: solution_Helmholtz} into \eqref{eq: Helmholtz}, assuming \cite{Levy2000Parabolic}
\begin{equation}
	\left|\frac{\partial^2 u}{\partial z^2}\right|\ll k_0 \left|\frac{\partial u}{\partial z}\right|,
\end{equation}
the standard PWE formulation is obtained:
\begin{equation}
\label{eq: parabolic_wave_equation}
\frac{\partial u }{\partial z} = \frac{1}{2j k_0}\left(\frac{\partial^2}{\partial x^2}+\frac{\partial^2}{\partial y^2}\right) u(x,y,z) .
\end{equation}
After using a split-step Fourier transform technique, the field distribution in each cross-section is updated from the field in the previous cross-section by \cite{Qin2024Efficient}
\begin{equation}
    \label{eq: pwe_Fourier}
    \hat{u}(k_x, k_y, z)=\mathcal{F}\{u(x,y,z);x,y\},
\end{equation}
\begin{equation}
\begin{split}
    \label{eq: update_Fourier}
    u(x, y, z+\Delta z)=\mathcal{F}^{-1}\{&\mathrm{exp}(-\tfrac{j(k_x^2)\Delta z}{2k_0})\mathrm{exp}(-\tfrac{j(k_y^2)\Delta z}{2k_0})\\
    &\hat{u}(k_x,k_y, z); k_x,k_y\}, 
\end{split}
\end{equation}
where $\mathcal{F}$ and $\mathcal{F}^{-1}$ denote the Fourier transform pairs in the transverse domain, $k_x$ and $k_y$ are spectral variables. When the wave reaches an obstacle, it is decomposed into forward and backward propagating components, as shown in Fig.~\ref{fig: two_way_PWE}. The forward propagating wave is predicted by \eqref{eq: update_Fourier}, and the backward propagating wave is updated by 
\begin{equation}
\begin{split}
    \label{eq: update_Fourier_backward}
    u(x, y, z-\Delta z)=\mathcal{F}^{-1}\{&\mathrm{exp}(\tfrac{j(k_x^2)\Delta z}{2k_0})\mathrm{exp}(\tfrac{j(k_y^2)\Delta z}{2k_0})\\
    &\hat{u}(k_x,k_y, z); k_x,k_y\},
\end{split}
\end{equation}
% Besides the forward wave propagation, the backward waves are also taken into account when the propagator reaches the obstacle. The waves are split into two components propagating in forward and backward directions while meeting obstacles, as illustrated by blue and red arrows in Fig.~\ref{fig: two_way_PWE}. Similar to the forward wave propagating update equation \eqref{eq: update_equation_Fourier}, the update equation for the backward propagating waves is
% \begin{equation}
% \label{eq: update_equation_Fourier_backward}
% u(x, y, z-\Delta z)=\mathcal{F}^{-1}\{C(k_x)C(k_y)\mathcal{F}\{u(x,y,z)\}\}
% \end{equation}

\section{Adaptive Polynomial Chaos Expansion}
\label{sec: adaptive_PCE}

\subsection{Overview}
\label{subsec: overview}

\par Polynomial chaos expansion provides a surrogate model for the dependence between a $N_d$-dimensional input random vector $\bm{\xi}$ and a quantity of interest (QoI) $ q(\bm{\xi})$ by expanding QoI onto an orthonormal polynomial basis \cite{Xiu2010Numerical}:
\begin{equation}
     \label{eq: PCE}
     q(\bm{\xi}) \approx \sum_{\bm{\alpha}\in\mathcal{A}}c_{\bm{\alpha}}\psi_{\bm{\alpha}}(\bm{\xi}),
\end{equation}
where $\psi_{\bm{\alpha}}(\bm{\xi})$ is the multivariate polynomial orthonormal with respect to the joint probability distribution of the input variables. To obtain a finite representation, the polynomial basis set $\mathcal{A}$ must be truncated. In standard PCE, this is commonly done using a maximum total-order $O_T$ criterion, as shown in Appendix \ref{sec: Static_PCE}. As summarized in Table~\ref{table: number_basis_total}, the number of polynomial basis functions increases rapidly with the maximum total order. Although alternative truncation strategies, such as hyperbolic truncation and maximum interaction order truncation, can reduce the basis size \cite{Blatman2011Adaptive, Blatman2008Sparse}, they still require user-defined truncation parameters. This motivates the use of an APCE method.

\par \rev{In this work, we develop a fixed number of samples APCE method to construct a compact and stable PCE model from a prescribed and limited set of samples. The overall procedure is illustrated in Fig.~\ref{fig:apce_workflow}. Starting from the available sample set, the polynomial basis is initialized and then adaptively extended using the basis extension algorithm adopted from \cite{Saturnino2019A}. This basis extension is driven by the variance contributions of the current polynomial terms. After each extension, the PCE coefficients are estimated by least squares regression, and the validation error is evaluated. The key feature of the proposed method is that the sample set is kept fixed during the entire adaptive procedure. Therefore, the adaptive procedure does not increase the number of samples as the basis grows, which makes the control of basis growth essential. To this end, a composite stopping criterion is introduced by combining the available sample size with the validation error trend. In this way, the proposed method selects a compact polynomial basis from the available samples and avoids excessive basis growth.}

\begin{figure}[!t]
\centering
\begin{tikzpicture}[
    node distance=8mm,
    >=latex,
    font=\small,
    process/.style={
        rectangle,
        draw,
        rounded corners=2pt,
        align=center,
        minimum width=0.82\linewidth,
        minimum height=8mm
    },
    decision/.style={
        diamond,
        draw,
        aspect=2.4,
        align=center,
        inner sep=1pt,
        minimum width=0.72\linewidth,
        minimum height=12mm
    },
    arrow/.style={->, thick}
]

% Nodes
\node[process] (samples) {
    \textbf{Fixed sample set} \\
    \footnotesize Input samples and corresponding QoIs
};

\node[process, below=of samples] (init) {
    \textbf{Initialize polynomial basis} \\
    \footnotesize Polynomial families from input distributions \cite{Xiu2002TheWiener} \\
    \footnotesize Initial basis set contains only the constant term
};

\node[process, below=of init] (extend) {
    \textbf{Implement basis extension} \\
    \footnotesize adopted from \cite{Saturnino2019A}
    
};

\node[process, below=of extend] (coeff) {
    \textbf{Estimate PCE coefficients} \\
    \footnotesize by least-squares regression \eqref{eq: OLS}    
};

\node[process, below=of coeff] (error) {
    \textbf{Calculate validation error}\\
    \footnotesize using LOOCV \eqref{eq: LOOCV_l} and \eqref{eq: LOOCV_multi}
};

\node[decision, below=of error] (stop) {
     \textbf{Stopping criterion satisfied?}\\
    \footnotesize Sample-to-basis ratio \& error trend \\
    \footnotesize proposed in this work
};

\node[process, below=of stop] (final) {
    \textbf{Final PCE model} \\
    \footnotesize Used for mean, STD and quantile estimation
};

% Arrows
\draw[arrow] (samples) -- (init);
\draw[arrow] (init) -- (extend);
\draw[arrow] (extend) -- (coeff);
\draw[arrow] (coeff) -- (error);
\draw[arrow] (error) -- (stop);

\draw[arrow] (stop) -- node[right, font=\footnotesize] {Yes} (final);

% Loop back
\draw[arrow] (stop.east) -- ++(1.0,0)
    node[midway, above, font=\footnotesize] {No}
    |- (extend.east);

\end{tikzpicture}
\caption{\rev{Workflow of the proposed fixed-sample APCE method. A fixed and limited set of PWE samples is used throughout the APCE process. The basis extension algorithm is adopted from~\cite{Saturnino2019A}, while the proposed stopping criterion controls basis growth according to the available sample size and the validation error convergence.}}
\label{fig:apce_workflow}
\end{figure}

\par \rev{Accordingly, the polynomial basis initialization step constructs the polynomial basis according to the probability distributions of the input variables. 
This selection is necessary for applying PCE, since the polynomial basis functions must be orthonormal with respect to the probability distribution of the input uncertain variables.
For the independent input variables considered in this work, the one-dimensional orthonormal polynomial family is selected according to the Wiener--Askey framework \cite{Xiu2002TheWiener}; for example, Hermite, Legendre, Laguerre, and Jacobi polynomials are commonly associated with Gaussian, uniform, gamma-type, and beta-type variables, respectively.
The multivariate basis functions are then formed by tensor products of the selected one-dimensional families.}

\par \rev{After this initialization step, the entire PCE construction in Fig.~\ref{fig:apce_workflow} is based on the selected polynomial family, including the basis extension, PCE coefficient estimation, validation error calculation, and final PCE model construction blocks.
}

\par Compared with standard and sparse PCE formulations summarized in Appendices \ref{sec: Static_PCE} and \ref{sec: Sparse_PCE}, the proposed APCE method selects the polynomial basis adaptively from the available samples, rather than prescribing a fixed total-order basis or relying on sparsity regularization. The adopted basis extension algorithm, coefficient calculation, validation error estimation, and composite stopping criterion are introduced in the following subsections.

% \subsection{Polynomial Family Selection}
% \par The orthonormality of the polynomial basis is ensured through univariate orthonormal polynomials $\psi_{\alpha_k^i}(\xi_i)$, which satisfy
% \begin{equation}
%     \label{eq: orthonormal}
%     \int_{\Xi_i}\psi_j^{(i)}(\xi_i)\psi_k^{(i)}(\xi_i)f_{\Xi_i}(\xi_i)\mathrm{d}\xi_i =
%     \begin{cases}
% \|\psi^{\,i}\|_{\Xi_i}^2, & \text{for } j = k,\\
% 0, & \text{else}.
% \end{cases}
% \end{equation}
% where the random variable $\xi_i$ is supported on the space $\Xi_i$ and under the probability distribution $f_{\Xi_i}(\xi_i)$. For the common probability distribution, the appropriate univariate polynomial families are selected according to the Wiener-Askey scheme \cite{Xiu2002TheWiener}. The multivariate polynomial basis functions $\Psi_{\bm{\alpha}}$ in \eqref{eq: PCE} are then constructed as a product of these corresponding univariate polynomials:
% \begin{equation}
%     \label{eq: mul_polynomial}
%     \Psi_{\bm{\alpha}} = \prod_{i=1}^{N_d}\psi_{\alpha_k^i}^{(i)}(\xi_i)
% \end{equation}

% \subsection{Sample Set Generation}

% \par To generate the sample set, one may use direct random sampling of the inputs. However, to prevent clustering and better coverage of the probability space, structured sampling, such as Latin Hypercube Sampling (LHS) \cite{Helton2003Latin} and optimal sampling, may also be employed \cite{Weise2022Comparison}.

\subsection{Basis Extension}
\label{subsec: basis_extension}

\par \rev{The basis extension in the proposed APCE method follows \cite{Saturnino2019A}. This algorithm uses two sets of multi-indices: the candidate set $\mathcal{A}_{\mathrm{cand}}$, which contains the multi-indices eligible for extension, and the old set $\mathcal{A}_{\mathrm{old}}$, which contains the multi-indices that have already been selected for extension.}

\par \rev{At each iteration, the multi-index with the largest variance contribution is selected from $\mathcal{A}_{\mathrm{cand}}$. Suppose that there are $N_q$ QoIs, and let $c_{\boldsymbol{\alpha}_k,i}$ denote the PCE coefficient associated with the basis function $\Psi_{\boldsymbol{\alpha}_k}(\boldsymbol{\xi})$ for the $i$-th QoI. The overall contribution of the multi-index $\boldsymbol{\alpha}_k$ is evaluated as
\begin{equation}
    \label{eq: normalized_coefficient}
    c_{\bm{\alpha}_k}^{\nu} = \sum_{i=1}^{N_q}(c_{{\bm{\alpha}_k},i})^2 .
\end{equation}
The multi-index with the largest $c^{\nu}_{\boldsymbol{\alpha}_k}$ is then used for basis extension.}

\par \rev{Let $\mathbf{e}_i$ denote the $i$-th unit multi-index vector, whose $i$-th component is one and all other components are zero. For a selected multi-index $\boldsymbol{\alpha}_{k}$, the basis extension algorithm first forms its forward neighboring multi-indices,
\begin{equation}
\boldsymbol{\beta}_i = \boldsymbol{\alpha}_{k} + \mathbf e_i,
\quad i=1,\ldots,N_d .
\end{equation}
A forward neighbor is not necessarily added to the candidate set. It is added only if it is admissible, namely if all its lower-order predecessor multi-indices have already been included in $\mathcal A_{\mathrm{old}}$. In other words, $\boldsymbol{\beta}_i$ is included only when
\begin{equation}
\label{eq: admissible_condition}
\boldsymbol{\beta}_i-\mathbf{e}_j \in \mathcal{A}_{\mathrm{old}},
\quad \forall j=1, \dots,N_d  \ \text{such that} \ \beta_{i.j}>0 
\end{equation}
where $\beta_{i,j}$ denotes the $j$-th component of the multi-index $\bm{\beta}_i$.  The multi-indices added to $\mathcal A_{\mathrm{cand}}$ are therefore admissible forward neighbors, rather than all forward neighbors. This admissibility condition ensures that whenever a multi-index is added, all of its lower-order predecessors have already been included. Hence, the basis extension does not introduce isolated higher-order terms without the corresponding lower-order predecessors, thereby preserving the downward closed structure of the polynomial basis, as in \cite{Saturnino2019A}.}

% \par At each iteration, multi-indices adjacent to the index with the largest variance contribution are added to the basis. Furthermore, to avoid overfitting of the PCE model, we impose a maximum polynomial order and exclude any polynomial whose order exceeds this limit. The anisotropic basis-extension procedure is summarized in Algorithm~\ref{alg:anisotropic_multilevel}.

% Two multi-indices are considered adjacent if they differ by one in exactly one component, i.e., if they can be written as $\bm{\alpha}_{k'} + \bm{e}_i$, where $d$ is the input dimension and $\bm{e}_i$ denotes the $i$-th canonical unit vector in $\mathbb{N}^d$. 

\subsection{Coefficient calculation and error evaluation}
\label{subsec: loocv}

\par The PCE coefficients are determined using a regression-based approach \cite{Xiu2010Numerical}. Let $\{\bm{\xi}^{(1)}, \dots, \bm{\xi}^{(N_s)} \}$ be a sample set of the input parameters, and let $\mathbf{q}=(q^{(1)},\dots,q^{(N_s)} )^T$ be the corresponding vector of model responses. Let us define the regression matrix $\bm{\Psi}$ with entries $\bm{\Psi}_{ij}=\psi_j(\bm{\xi}^{(i)})$. According to \eqref{eq: PCE}, the coefficients $\mathbf{c}$ can be calculated by least squares regression:
\begin{equation}
    \label{eq: OLS}
    \hat{\mathbf{c}}=\arg\min \|\bm{\Psi}\mathbf{c}-\mathbf{q}\|_2^2
\end{equation}
Owing to the basis extension algorithm, the resulting PCE basis is sufficiently compact that no additional $\ell_1$ or $\ell_2$ regularization is required in the coefficient estimation step.

% Therefore, the PCE coefficients are calculated directly from ordinary least squares regression \eqref{eq: OLS}, whose closed-form solution is given by the Moore-Penrose pseudoinverse:
% \begin{equation}
%     \label{eq: Moore_Penrose}
%     \hat{\mathbf{u}} = (\bm{\Psi}^T \bm{\Psi})^{-1}\bm{\Psi}^T\bm{q}
% \end{equation}

\par\rev{
The validation error at each iteration is evaluated using the leave-one-out cross-validation (LOOCV) error. Since the PCE coefficients in the proposed APCE method are estimated by least-squares regression, the analytical LOOCV formulation can be directly applied \cite{BlatmanThesis}. Let $h_i$ be the $i$-th component of the vector given by
\begin{equation}
    \label{eq: def_h}
    \mathbf{h} = \mathrm{diag}(\bm{\Psi}(\bm{\Psi}^T \bm{\Psi})^{-1}\bm{\Psi}^T) ,
\end{equation}
For the $l$-th QoI, the LOOCV error is computed as
\begin{equation}
    \label{eq: LOOCV_l}
    \varepsilon^{\mathrm{LOO}}_l =
    \frac{
    \sum_{i=1}^{N_s}
    \left(
    \frac{
    q_l(\bm{\xi}^{(i)}) -
    \sum_{\bm{\alpha}\in\mathcal{A}}
    c_{\bm{\alpha},l}\psi_{\bm{\alpha}}(\bm{\xi}^{(i)})
    }{
    1-h_i
    }
    \right)^2
    }{
    \sum_{i=1}^{N_s}
    \left(
    q_l(\bm{\xi}^{(i)})-\bar q_l
    \right)^2
    },
\end{equation}
where
\begin{equation}
    \bar q_l=\frac{1}{N_s}\sum_{i=1}^{N_s}q_l(\bm{\xi}^{(i)}).
\end{equation}
For multi-output QoIs, the overall validation error is obtained by averaging over all output components:
\begin{equation}
    \label{eq: LOOCV_multi}
    \varepsilon_v =
    \frac{1}{N_q}
    \sum_{l=1}^{N_q}
    \varepsilon^{\mathrm{LOO}}_l .
\end{equation}
}

% The PCE coefficients are commonly determined using a regression-based approach. Given a sample set $\{\bm{\xi}^{(1)}, \dots, \bm{\xi}^{(N_s)} \}$ of the input parameters, and the corresponding model evaluations $\mathbf{Q}=(q^{(1)},\dots,q^{(N_s)} )^T$. Let us define the regression matrix $\bm{\Psi}$ with entries $\bm{\Psi}_{ij}=\Psi_{\bm{\alpha}_j}(\bm{\xi}^{(i)})$. According to \eqref{eq: PCE}, the coefficients $\mathbf{u}$ can be calculated by ordinary least squares:
% \begin{equation}
%     \label{eq: OLS}
%     \hat{\mathbf{u}}=\arg\min \|\bm{\Psi}\mathbf{u}-\mathbf{Q}\|_2^2,
% \end{equation}

\subsection{Composite Stopping Criterion}
\label{subsec: termination_criteria}

\par \rev{A suitable stopping criterion is essential for achieving good approximation performance, especially under limited sampling budgets. Existing APCE methods typically rely on a single stopping rule, such as a hard threshold on the properties of the regression matrix \cite{Loukrezis2025Multivariate, Hampton2018Basis} or on the approximated error \cite{Wegert2025Conductivity, Saturnino2019A}. Under a limited sampling budget, however, a single-threshold rule is fragile: it may terminate the extension before the validation error becomes a reliable indicator, or, conversely, continue extending the basis until the least squares system becomes ill-conditioned. To address this issue, we propose a composite heuristic stopping criterion that simultaneously controls (i) premature termination, (ii) regression conditioning, and (iii) overfitting. Specifically, the basis extension is stopped according to the following rules:}
\begin{itemize}
    \item the number of polynomial terms is required to exceed $0.25$ times the number of available samples, to avoid premature termination.
    \item to ensure the accuracy and stability of the regression, the algorithm is forced to stop once the number of polynomial terms exceeds $0.5$ times the number of samples;
    \item if the minimum LOOCV $\varepsilon_v$ does not decrease for consecutive iterations, the algorithm stops early to avoid overfitting;
    % \item if the validation error reaches the prescribed target threshold, the algorithm terminates.
\end{itemize}
The lower and upper bounds of the polynomial terms are motivated by the empirical rule of thumb that, when the number of samples is about four times the number of unknown coefficients, the ordinary least-squares estimate in \eqref{eq: OLS} is already sufficiently accurate; and that accurate and stable ordinary least-squares regression typically requires at least twice as many samples as polynomial coefficients \cite{Luthen2021sparse}.

\section{Examples in Realistic Irregular Terrains}
\label{sec: examples}

\subsection{Overview}
\label{subsec: overview}

\par
\begin{figure}[!htbp]
	% \captionsetup{font=footnotesize}
	\centering
    \includegraphics[width=\columnwidth]{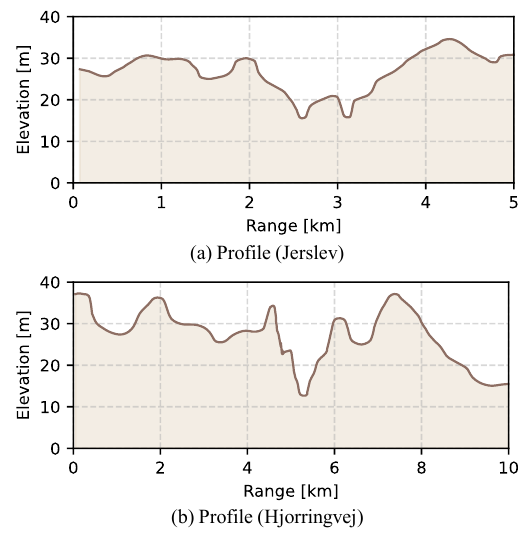}
    \caption{Jerslev and Hjorringvej terrain profiles used to validate the proposed APCE method.}
	\label{fig: terrain_profiles}
\end{figure}

 We validate the proposed framework over two realistic terrain profiles in northern Denmark, namely Jerslev and Hjorringvej \cite{Hviid1995Terrain}, shown in Fig. \ref{fig: terrain_profiles}. In all cases, the ground is modeled as a lossy dielectric with relative permittivity $\varepsilon_r = 4.5$ and loss tangent $\tan\delta = 0.07$. For the two-way PWE simulations, the spatial steps are set to $\Delta x=50$~m and $\Delta y=0.5$~m, based on \cite{Qin2024Comparative}. A unit-amplitude Gaussian source is used, and the total propagation distances are $5$~km and $10$~km for the Jerslev and Hjorringvej profiles, respectively.

\par
\begin{figure}[!htbp]
    \centering
    \includegraphics[width = \columnwidth]{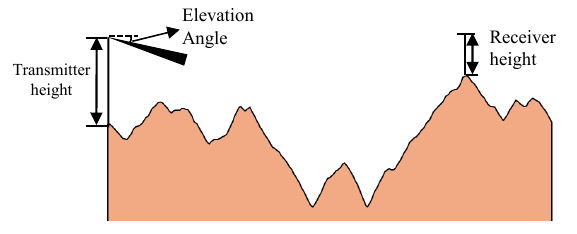}
    \caption{Geometry of the antenna configuration and definition of the uncertain geometric parameters: transmitter height, elevation angle, and receiver height.}
    \label{fig: geometry_parameters}
\end{figure}

 We consider the uncertainties from two sources. The first group consists of geometric parameters, including transmitter height,  elevation angle, and receiver height, as illustrated in Fig.~\ref{fig: geometry_parameters}. The second group includes internal transmitter parameters, such as the operating frequency and the antenna beamwidth. All uncertain parameters are modeled as Beta random variables with shape parameters $(3,3)$ and bounded support, as summarized in Table~\ref{table: uncertain_parameters}. This choice yields a Gaussian-like distribution while enforcing strict physical bounds on each parameter, which is particularly appropriate for modeling installation tolerances and equipment variations that cannot exceed prescribed limits~\cite{Saturnino2019A, Gupta2004Handbook}.

\begin{table}[!htbp]
\centering
\renewcommand\arraystretch{1.2}
\caption{Uncertain Antenna Configuration Parameters and Their Bounds. Their Probability $\mathrm{Beta}(3,3)$ Denotes a Beta Distribution with Shape Parameters $3$ and $3$}
\label{table: uncertain_parameters}
\begin{tabular}{@{}c@{\hspace{2pt}}c@{\hspace{2pt}}c@{}c@{}c@{}c@{}c@{}c@{}}
    \toprule
    \textbf{Parameter} &  \textbf{Unit} &\textbf{Distribution} & \textbf{Interval} & \textbf{Reference} \\
    \midrule
    \makecell[c]{Transmitter \\height}  & m            & $\mathrm{Beta}(3,3)$ & $[9, 13]$                    & \cite{Hviid1995Terrain} \\
    \makecell[c]{Receiver \\height}     & m            & $\mathrm{Beta}(3,3)$ & $[1, 4]$                      & \cite{Hviid1995Terrain} \\
    \makecell[c]{Elevation \\angle}     & \si{\degree} & $\mathrm{Beta}(3,3)$ & $[-3, 3]$                  & \cite{IEEEstd1492021, ITU_R_F1336} \\
    Beamwidth                           & \si{\degree} & $\mathrm{Beta}(3,3)$ & $[4, 12]$              & \cite{IEEEstd1492021, ITU_R_F1336} \\ 
    \multirow{2}*{\makecell[c]{Frequency}} & \multirow{2}*{\makecell[c]{MHz}} & \multirow{2}*{\makecell[c]{$\mathrm{Beta}(3,3)$}} & $[410, 460]$ (Jerslev) &  \multirow{2}*{\makecell[c]{\cite{IEEEstd1492021, ITU_R_F1336, Hviid1995Terrain}}}\\
                                         &                                    &                                                   & $[920, 1020]$ (Hjorringvej)  & \\        
    \bottomrule
\end{tabular}
\label{table: uncertain_parameters}
\end{table}

\par The QoIs are the path loss along the propagation range, which is proportional to the electric field component parallel to the receiving antenna polarization. Let $r_i$, $i=1, \dots, N_r$ denote the equidistant locations, the path loss at position $r_i$ is calculated as \cite{Apaydin2017Radio}
\begin{equation}
    \label{eq: received_power}
    q(r_i)=20\log_{10}|E_{\|}(r_i,h_{\mathrm{Rx}})|+C,
\end{equation}
where $E_{\|}(r_i,h_{\mathrm{Rx}})$ denotes the electric field component parallel to the receiving antenna polarization at receiver location $(r_i,h_{\mathrm{Rx}})$, and $C$ denotes a distance-independent calibration offset accounting for source normalization and antenna-related factors. Collecting these received powers into a vector, we define $\mathbf{q}=[q_1, \dots, q_{N_r}]^{\mathsf{T}}$ with $q_i = q(r_i)$. From the PCE model, we estimate at each spatial location the mean $\mathbb{E}(q_i)$, the standard deviation (STD) $\mathbb{S}(q_i)$, the 5th percentile $\mathbb{Q}^5(q_i)$, and the 95th percentile $\mathbb{Q}^{95}(q_i)$, which we assemble into the vectors $\mathbb{E}(\mathbf{q})$, $e_r(\mathbb{S}(\mathbf{q}))$, $\mathbb{Q}^5(\mathbf{q})$, and $\mathbb{Q}^{95}(\mathbf{q})$, respectively. The mean characterizes the expected propagation behavior, while the 5th and 95th percentiles define a $90\%$ reliability interval along the propagation range. 
\rev{To assess the accuracy of various methods, the relative error of a statistical quantity is defined as
\begin{equation}
\label{eq: err_stat}
e_r(\mathcal{T}(\mathbf{q}))
:= 
\frac{
\sqrt{\sum_{i=1}^{N_r}[\mathcal{T}(q_i)
-
\mathcal{T}_{\mathrm{ref}}(q_i)]^2}}
{\sqrt{\sum_{i=1}^{N_r}[\mathcal{T}_{\mathrm{ref}}(q_i)]^2}},
\end{equation}
where $\mathcal{T}\in\left\{\mathbb{E},\mathbb{S},\mathbb{Q}^{5},\mathbb{Q}^{95}\right\}$. Here,  $\mathbb{E}(\cdot)$, $\mathbb{S}(\cdot)$, $\mathbb{Q}^5(\cdot)$, and $\mathbb{Q}^{95}(\cdot)$ are the mean, STD, $5$-th percentile, and $95$-th estimators, whereas $\mathcal{T}_{\mathrm{ref}}(\cdot)$ are the reference estimates computed from $10^5$ MC simulations. }
The details about these estimators are shown in Appendix~\ref{sec: postprocessing}.

\subsection{Convergence Analysis}
\label{subsec: CA}

\par
\begin{figure}[!htbp]
	% \captionsetup{font=footnotesize}
	\centering
    \includegraphics[width=0.95\columnwidth]{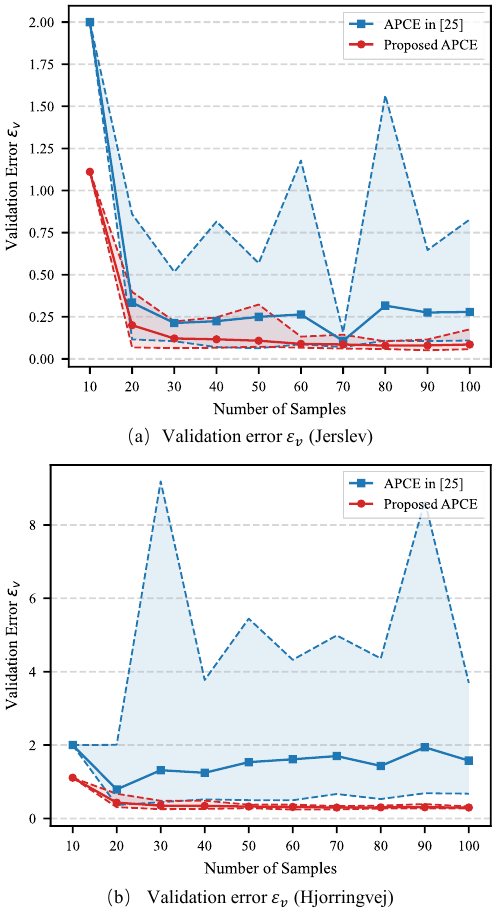}
     \caption{\rev{Validation error $\varepsilon_v$ [defined in \eqref{eq: LOOCV_multi}] of the PCE model versus the number of training samples $N_s$ for the Jerslev and Hjorringvej terrain profiles. The solid lines represent the mean validation error, while the shaded bands bounded by dashed lines indicate the minimum--maximum range across the trials.}}
    \label{fig: loocv}
\end{figure}

 The proposed APCE method automatically selects the polynomial basis functions used in the PCE model. Before estimating the path-loss statistics, we first investigate the convergence behavior of the proposed APCE method with respect to the number of training samples. To this end, we analyze the LOOCV error, as defined in \eqref{eq: LOOCV_multi}, for different sample sizes. Ten sampling levels are considered, $N_s = 10, 20, 30, 40, 50, 60, 70, 80, 90, 100$. For each sample level $N_s$, thirty independent training sets are generated, and corresponding PCE models are constructed. We further compare the proposed APCE method with the method in \cite{Saturnino2019A}, which uses the same basis extension algorithm. The resulting distributions of the validation error are summarized in Fig.~\ref{fig: loocv}.

\par \rev{
As shown in Fig.~\ref{fig: loocv}, the validation error of the proposed APCE decreases rapidly when the number of samples increases from  $N_s=10$ to $N_s=30$. After the initial reduction, the error remains at a low level with only small fluctuations as $N_s$ further increases. This behavior indicates that the proposed method can construct a stable PCE model once a moderate number of training samples is available. The error bands of the proposed APCE are also narrower for most sampling levels, showing limited trial-to-trial variability over the 30 independent training sets. Compared with the APCE method of~\cite{Saturnino2019A}, the proposed method yields lower mean LOOCV errors and narrower minimum--maximum error bands, indicating more stable PCE model construction.}

\par To further analyze the adaptivity of the proposed APCE, we report in Tables~\ref{table: N_poly_different_sample_size_Jerslev} and~\ref{table: N_poly_different_sample_size_Hjorringvej} the number of selected polynomials, the maximum polynomial order, and the maximum interaction order of the polynomial bases constructed by the proposed APCE method. \rev{The maximum polynomial order reported in Tables~\ref{table: N_poly_different_sample_size_Jerslev} and~\ref{table: N_poly_different_sample_size_Hjorringvej} is defined as the highest order of any individual polynomial basis function selected in the final basis, whereas the maximum interaction order is defined as the largest number of distinct variables involved in a single multivariate basis term. In all APCE simulations, the maximum admissible polynomial order is set to five. This value is not used to prescribe a complete fifth-order PCE basis, but only to restrict the candidate search space during adaptive basis extension. The basis functions are added progressively according to their variance contributions, and the final polynomial basis is determined by the selected terms and the stopping criterion.} From Tables~\ref{table: N_poly_different_sample_size_Jerslev} and~\ref{table: N_poly_different_sample_size_Hjorringvej}, it can be observed that the proposed method automatically adjusts the number of selected basis functions as the sample size increases, while effectively suppressing high-order interaction terms. In particular, already at $N_s \approx 30$, the selected bases remain compact and are dominated by low interaction orders, with higher-order interactions appearing only occasionally as $N_s$ becomes larger. This behavior suggests that the dominant contributions to the response can be captured using relatively low-order and low-interaction polynomials, which improves robustness and reduces the risk of overfitting. Since both the validation error and the basis information exhibit diminishing returns beyond $N_s \approx 30$, we adopt $N_s=30$ as a cost-effective sampling budget in the subsequent UQ study and use larger $N_s$ only for additional robustness comparisons. 

\begin{table}[!htbp]
\centering
\renewcommand\arraystretch{1.2}
\caption{Number of basis functions, maximum polynomial order, and maximum interaction order selected under different sample sizes (Jerslev). Polynomial order and interaction order of $0$ indicates that only the constant term is selected.}
\begin{tabular}{c c c c}
    \toprule
    \makecell[c]{ \textbf{Number of} \\ \textbf{Samples} }  & \makecell[c]{ \textbf{Number of} \\ \textbf{Basis}\\ \textbf{Functions} }  &\makecell[c]{ \textbf{Maximum} \\ \textbf{Polynomial} \\\textbf{Order} }  &\makecell[c]{ \textbf{Maximum} \\ \textbf{Interaction} \\ \textbf{Order} } \\
    \midrule
    10  & 1  & 0 & 0 \\
    20  & 9  & 2 & 2 \\
    30  & 13 & 3 & 2 \\
    40  & 18 & 3 & 2 \\
    50  & 23 & 3 & 2 \\
    60  & 21 & 4 & 2 \\
    70  & 22 & 5 & 2 \\
    80  & 30 & 5 & 2 \\
    90  & 29 & 5 & 3 \\
    100 & 32 & 5  & 2\\
    \bottomrule
\end{tabular}
\label{table: N_poly_different_sample_size_Jerslev}
\end{table}

\begin{table}[!htbp]
\centering
\renewcommand\arraystretch{1.2}
\caption{Number of basis functions, maximum polynomial order, and maximum interaction order selected under different sample sizes (Hjorringvej). Polynomial order and interaction order of $0$ indicates that only the constant term is selected.}
\begin{tabular}{c c c c}
    \toprule
    \makecell[c]{ \textbf{Number of} \\ \textbf{Samples} }  & \makecell[c]{ \textbf{Number of} \\ \textbf{Basis}\\ \textbf{Functions} }  &\makecell[c]{ \textbf{Maximum} \\ \textbf{Polynomial} \\\textbf{Order} }  &\makecell[c]{ \textbf{Maximum} \\ \textbf{Interaction} \\ \textbf{Order} } \\
    \midrule
    10  & 1  & 0 & 0 \\
    20  & 9  & 2 & 2 \\
    30  & 14 & 4 & 2 \\
    40  & 19 & 3 & 2 \\
    50  & 21 & 4 & 2 \\
    60  & 22 & 4 & 2 \\
    70  & 26 & 3 & 2 \\
    80  & 28 & 4 & 2 \\
    90  & 31 & 3 & 3 \\
    100 & 34 & 5 & 2\\
    \bottomrule
\end{tabular}
\label{table: N_poly_different_sample_size_Hjorringvej}
\end{table} 

% \par In this convergence study, we deliberately set a very strict target error of $10^{-3}$, which is not reached in the plotted curves in Fig.~\ref{fig: loocv}. This is by design: once the error stops decreasing significantly, the composite stopping criterion triggers an early stop, so that basis extension is halted automatically before the prescribed target error is attained, thereby avoiding overfitting and unnecessary growth of the polynomial basis. Since both the validation error and the basis information exhibit diminishing returns beyond $N_s \approx 30$, we adopt $N_s=30$ as a cost-effective sampling budget in the subsequent UQ study and use larger $N_s$ only for additional robustness comparisons.

% \subsection{Mean and Std Estimation}
\subsection{Uncertainty Quantification}
\label{subsec: UQ}

% \subsubsection{Accuracy for a fixed sampling number}

\par
\begin{figure*}[!tbp]
    \centering
    \includegraphics[width=0.95\linewidth]{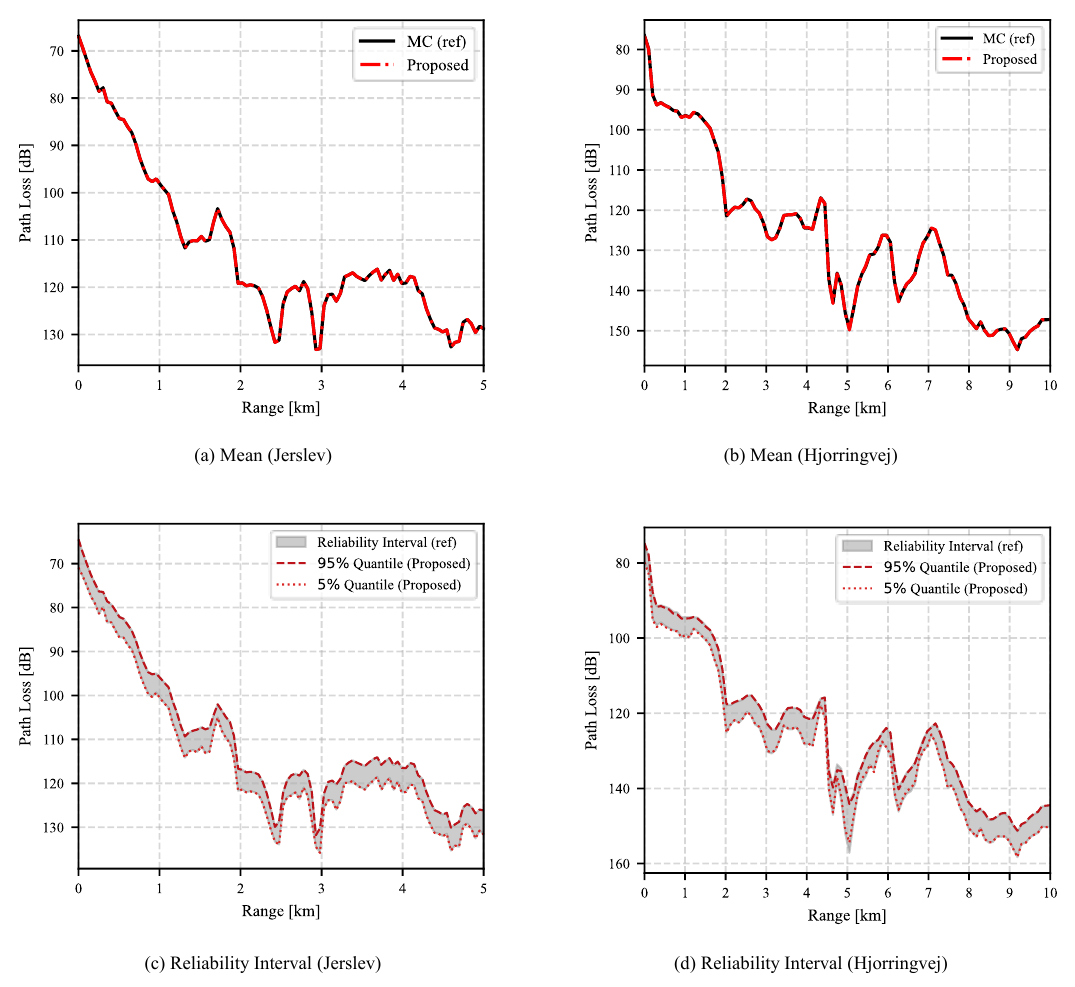}
    \caption{Mean and reliability interval (5\%--95\% quantiles) of the path loss versus range for the Jerslev and Hjorringvej terrain profiles. The proposed APCE method with $30$ PWE simulations is validated against the Monte Carlo (MC) reference using $10^5$ simulations.}
	\label{fig: mean_interval}
\end{figure*}

 With $N_s = 30$ samples, we construct a PCE model using the proposed APCE method. Fig.~\ref{fig: mean_interval} shows the mean path loss and the reliability interval (defined by the $5\%$ and $95\%$ quantiles) along the propagation range for the two terrains, estimated by the proposed APCE and MC reference. For both Jerslev and Hjorringvej, the APCE mean curve closely agrees with MC reference results over the entire distance, including regions with pronounced local fading-like fluctuations. The reliability interval is also well reproduced. In Fig.~\ref{fig: mean_interval}, the $5$-th and $95$-th percentiles estimated by the APCE closely follow the edges of the MC reliability interval, indicating that the proposed APCE method captures not only the expected path loss but also the spread of path loss. 

\par To further compare the proposed approach against other PCE methods, namely the standard PCE described in Appendix~\ref{sec: Static_PCE} and the sparse PCE constructed using the least angle regression (LARS) algorithm shown in Appendix~\ref{sec: Sparse_PCE}. Table~\ref{table: Error_Jerslev_Hjorringvej} reports the CPU time for constructing the PCE model and the minimum and maximum relative errors over thirty independent training sets for the mean, STD, 5th percentile, and 95th percentile. All methods use the same training sets ($N_s=30$). \rev{The reported CPU time only accounts for the construction of the PCE model and excludes the deterministic PWE simulations. For a fixed sampling budget, all three PCE-based methods use the same training samples and therefore require the same number of PWE model evaluations.}

\par For the mean path loss, all PCE methods achieve sub-percent accuracy, suggesting that the average trend can be captured with limited samples.
% Nonetheless, APCE consistently yields the smallest error range, with the most notable improvement on Hjorringvej, where the maximum mean error decreases from $0.51\%$ (standard PCE) and $0.68\%$ (sparse PCE) to $0.27\%$.
\rev{The STD is more challenging to estimate than the mean and percentile bounds, as reflected by the larger relative errors in Table~\ref{table: Error_Jerslev_Hjorringvej}. Nevertheless, the proposed APCE substantially reduces the $e_r(\mathbb{S}(\mathbf{q}))$ compared with the standard and sparse PCE methods under the same sampling budget. The improvement is also observed for the reliability interval, where APCE yields smaller errors for both the 5th and 95th percentiles. Although the STD errors are relatively larger than those of the percentile bounds, the resulting reliability interval remains accurately captured. This indicates that, from the perspective of the uncertainty bounds considered here, the STD estimation error remains acceptable under the fixed sampling budget.
Overall, these results indicate that APCE is particularly effective in capturing the mean behavior and uncertainty bounds under a limited training budget, while also improving the estimation of the distribution spread compared with the other PCE methods.}

\par \rev{In terms of computational cost, the construction time of the proposed APCE is close to that of standard PCE and is much smaller than that of sparse PCE based on LARS. Compared with the deterministic simulation cost, the time for model construction of all three PCE methods remains acceptable. Overall, the proposed APCE achieves improved accuracy and robustness with only a negligible additional construction cost relative to standard PCE and a substantially lower construction cost than sparse PCE.}

\begin{table*}[!htbp]
    \centering
    \renewcommand\arraystretch{1.3}
    \caption{\rev{CPU time for constructing the PCE model and Minimum and maximum relative errors of the statistical quantities \eqref{eq: err_stat} of the path loss over 30 independent training sets ($N_s=30$).}}
    \fontsize{8pt}{9pt}\selectfont 
    % \begin{tabular}{@c@{\hspace{3pt}}c@{}c@{\hspace{3pt}}c@{\hspace{2pt}}c@{\hspace{3pt}}c@{\hspace{2pt}}c@{}}
    \begin{tabular}{ c c c c c c c c c c c}
    \toprule
    \textbf{Terrain} & \textbf{Method} & \makecell[c]{\rev{\textbf{CPU}}\\ \rev{\textbf{Time (s)} }}  & \multicolumn{2}{c}{\makecell[c]{$e_r(\mathbb{E}(\mathbf{q}))$\\ \textbf{Min.} \quad \textbf{Max.} } } & \multicolumn{2}{c}{\makecell[c]{\rev{$e_r(\mathbb{S}(\mathbf{q}))$}\\ \textbf{Min.} \quad \textbf{Max.} } } & \multicolumn{2}{c}{\makecell[c]{$e_r(\mathbb{Q}^5(\mathbf{q}))$\\ \textbf{Min.} \quad \quad \textbf{Max.}  } } & \multicolumn{2}{c}{\makecell[c]{$e_r(\mathbb{Q}^{95}(\mathbf{q}))$\\ \textbf{Min.} \quad \quad \textbf{Max.}  } }\\

    \midrule
    \multirow{3}*{\makecell[c]{Jerslev}} & Standard PCE  & \rev{$0.08$}      & $0.02\%$      & $0.64\%$      & \rev{$2.66\%$}      & \rev{$69.35\%$}     & $0.07\%$      & $1.22\%$      & $0.15\%$      & $1.96\%$     \\ 
                                         & Sparse PCE    & \rev{$11.58$}     & $0.02\%$      & $0.77\%$      & \rev{$7.70\%$}      & \rev{$59.58\%$}     & $0.12\%$      & $1.01\%$      & $0.23\%$      & $1.82\%$     \\ 
                                         & Proposed APCE & \rev{$\bm{0.09}$} & $\bm{0.01\%}$ & $\bm{0.67\%}$ & \rev{$\bm{1.99\%}$} & \rev{$\bm{16.39\%}$} & $\bm{0.05\%}$ & $\bm{0.75\%}$ & $\bm{0.04\%}$ & $\bm{0.75\%}$ \\
    \midrule                                     
    \multirow{3}*{\makecell[c]{Hjorringvej}} & Standard PCE  & \rev{$0.08$}      & $0.17\%$      & $0.51\%$      & \rev{$15.97\%$}     & \rev{$105.75\%$}     & $0.35\%$       & $2.59\%$      & $0.48\%$      & $2.54\%$     \\ 
                                             & Sparse PCE    & \rev{$11.82$}     & $0.17\%$      & $0.68\%$      & \rev{$23.17\%$}     & \rev{$170.30\%$}     & $0.40\%$       & $3.34\%$      & $0.69\%$      & $4.14\%$     \\ 
                                             & Proposed APCE & \rev{$\bm{0.09}$} & $\bm{0.06\%}$ & $\bm{0.27\%}$ & \rev{$\bm{8.98\%}$} & \rev{$\bm{29.34\%}$} & $\bm{0.33\%}$ & $\bm{0.79\%}$ & $\bm{0.26\%}$ & $\bm{0.66\%}$ \\
    \bottomrule
    \end{tabular}
    \label{table: Error_Jerslev_Hjorringvej}
\end{table*}

\begin{figure*}[!htbp]
	% \captionsetup{font=footnotesize}
	\centering
    \includegraphics[width=0.95\linewidth]{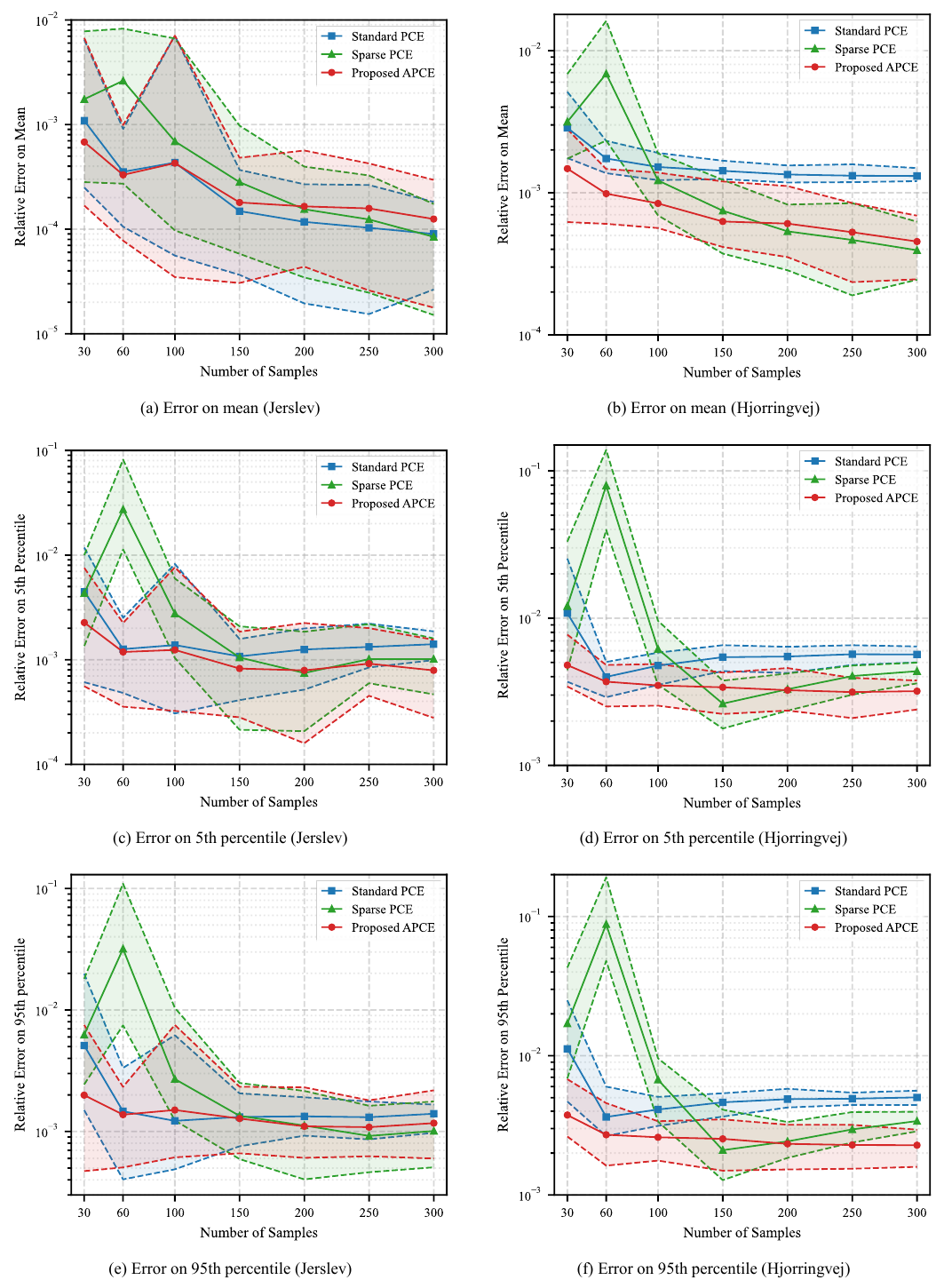}
    \caption{Relative error statistics versus the number of training samples for the Jerslev and Hjorringvej terrain profiles. Solid lines show the mean relative errors (over 30 independent training sets), and shaded bands (bounded by dashed lines) indicate the minimum--maximum error band across the trials. Results are shown for the mean (a)--(b), the 5th percentile (c)--(d), and the 95th percentile (e)--(f), comparing the Standard PCE, sparse PCE, and the proposed APCE.}
	\label{fig: mean_Q5_Q95}
\end{figure*}

\par To further assess the robustness of the proposed method with respect to the training samples, we investigate how the relative error \eqref{eq: err_stat} evolves as the number of training samples increases. Specifically, we consider $N_s \in \{30, 60, 100, 150, 200, 250, 300\}$. For each $N_s$, thirty independent training sets are generated and used to construct the corresponding PCE models. The relative errors of the estimated mean, 5th percentile, and 95th percentile of the path loss are then computed with respect to the MC reference. Fig.~\ref{fig: mean_Q5_Q95} summarizes the results, where the solid lines show the average (over the 30 trials) of the relative error and the shaded bands indicate the minimum--maximum of errors across the trials.

\par As shown in Fig.~\ref{fig: mean_Q5_Q95}(a)–(b), all methods yield small mean errors, and the average error generally decreases as $N_s$ increases. For Jerslev, the Standard PCE error also exhibits a clear downward trend with increasing $N_s$, reaching a comparable level to the proposed APCE at large sample sizes. Across most of the sampling range, APCE achieves the lowest or near-lowest average mean error, while maintaining consistently narrow error bands, indicating strong robustness to the choice of training sets. In contrast, the sparse PCE is noticeably less stable for small $N_s$ (e.g., around $N_s=60$), where both the average error and the spread increase markedly, suggesting a higher sensitivity to the particular training realization in the undersampled regime.

\par The differences become more pronounced for the quantile estimation that defines the reliability interval. In Fig.~\ref{fig: mean_Q5_Q95}(c)–(f), APCE maintains low average errors for both the 5th and 95th percentiles and shows a markedly narrower error range, demonstrating improved robustness to the training-set selection. By comparison, the sparse PCE again exhibits large variability at small $N_s$, with occasional error spikes and wide bands, while the standard PCE tends to level off at a higher error floor and does not consistently reduce its error spread as $N_s$ grows.

\section{Conclusion}
\label{sec: conclusion}

\par This paper has presented an efficient uncertainty quantification method for radio wave propagation over irregular terrains by combining the two-way parabolic wave equation method with an adaptive polynomial chaos expansion (APCE). The proposed APCE constructs a compact PCE model from a prescribed and limited set of training samples. The polynomial basis is extended according to variance contributions, while the basis growth is controlled by the LOOCV error and the available sample size. In this way, the method avoids prescribing a complete total-order basis and does not rely on strong sparsity assumptions.

\par Two realistic terrain profiles, Jerslev and Hjorringvej, have been investigated. For both cases, the proposed APCE provides mean, $5\%$ quantile, and $95\%$ quantile estimates of the path loss that closely match MC reference results while requiring only a small number of PWE simulations. In addition to the mean and percentile bounds, the standard deviation of the path loss has also been evaluated. The results show that the standard deviation is more difficult to estimate than the mean and the percentile bounds; nevertheless, the proposed APCE clearly reduces the error of STD estimation compared with both the standard PCE and the sparse PCE. This indicates that the adaptive basis selection improves not only the prediction of the uncertainty bounds but also the estimation of the distribution spread under a fixed sampling budget.

\par \rev{The present study focuses on the UQ of radio wave propagation over realistic irregular terrains with uncertain antenna configurations. The numerical examples involve five uncertain input parameters and two realistic terrain profiles with different frequency ranges. Although the proposed APCE reduces the number of retained polynomial terms by adaptive basis extension, problems with larger stochastic dimensions have not been tested in this work. Extending APCE to higher-dimensional UQ settings is an important and promising topic, and incorporating techniques such as dimensionality reduction may further broaden the applicability of the method. In addition, lower-frequency scenarios with longer wavelengths may exhibit different terrain-interaction and diffraction characteristics. Extending the proposed APCE method to broader frequency regimes, additional propagation environments, and higher-dimensional UQ settings will be considered in future work.
}

\appendices
\renewcommand{\thetable}{\thesection-\Roman{table}}
\setcounter{table}{0}

\section{Standard Polynomial Chaos Expansion}
\label{sec: Static_PCE}

\par Standard PCE refers to a PCE constructed using a polynomial basis set $\mathcal{A}$, which is truncated using the maximum total-order criterion with order $O_T$. The expansion coefficients are estimated via ordinary least-squares regression \eqref{eq: OLS}. The corresponding multi-indices set is:
\begin{equation}
    \label{eq: max_total_order}
    \Lambda_{O_{T_{max}}}^{N_d} = \{\bm{\alpha}\in \mathbb{N}^{N_d}: \sum_{i=1}^{N_d}\alpha_i\leq O_{T_{max}}\}
\end{equation}
whose cardinality is
\begin{equation}
    \label{eq: card_max_total}
    \mathrm{card}( \Lambda_{O_{T}}^{N_d}) = \binom{N_d +O_{T}}{N_d} =\frac{(N_d+O_{T})!}{N_d!O_{T}!} ,
\end{equation}
where $N_d$ denotes the number of uncertain input parameters. For the five-dimensional problem considered in this paper ($N_d = 5$), the resulting number of polynomial basis functions for different choices of the maximum total order is summarized in Table~\ref{table: number_basis_total}.
\begin{table}[!htbp]
\centering
\renewcommand\arraystretch{1.2}
\caption{Number of Polynomial Basis Functions for Standard PCE ($N_d = 5$)}
\begin{tabular}{c c c c c c c c }
    \toprule
    \textbf{Maximum Total Order}   & 1 & 2  & 3  & 4   & 5   & 6   & 7 \\
    \midrule
    \textbf{Number of Polynomials} & 6 & 21 & 56 & 126 & 252 & 462 & 792 \\
    \bottomrule
\end{tabular}
\label{table: number_basis_total}
\end{table}

\section{Sparse Polynomial Chaos Expansion}
\label{sec: Sparse_PCE}

\par To promote sparsity in the PCE representation and enable the estimation of the expansion coefficients from a limited number of samples, an $\ell_1$-regularized regression formulation is adopted:
\begin{equation}
    \label{eq: sparse_regression}
    \hat{\mathbf{c}}
    = \arg\min_{\mathbf{c}}
    \big\|\bm{\Psi}\mathbf{c} - \mathbf{q}\big\|_2^2
    + \lambda \|\mathbf{c}\|_1,
\end{equation}
which leads to the so-called sparse PCE \cite{Luthen2021sparse}. The optimization problem \eqref{eq: sparse_regression} can be solved by various sparse-regression algorithms, such as LARS \cite{Blatman2011Adaptive}.

\section{Postprocessing}
\label{sec: postprocessing}

\par Once the PCE model is constructed, the mean of the QoI can be calculated analytically from the PCE coefficients as
\begin{equation}
     \label{eq: get_mean}
     \mathbb{E}(q) = c_{\bm{\alpha}_0}
\end{equation}
where $\bm{\alpha}_0 = [0,0,\cdots, 0]$. The variance is obtained from all coefficients except for the one corresponding to $\bm{\alpha}_0$:
\begin{equation}
     \label{eq: get_var}
     \mathbb{V}(q)= \sum_{\bm{\alpha}\in \mathcal{A}\backslash \bm{\alpha}_0}(c_{\bm{\alpha}})^2
\end{equation}
and the standard deviation follows directly:
\begin{equation}
     \label{eq: get_std}
     \mathbb{S}(q) = \sqrt{\mathbb{V}(q)} = \sqrt{\sum_{\bm{\alpha}\in \mathcal{A}\backslash \bm{\alpha}_0}(c_{\bm{\alpha}})^2}
\end{equation}
Percentile estimates cannot be obtained in closed form from the coefficients alone. Instead, they are efficiently computed by sampling the PCE model using a Monte Carlo procedure. Since the evaluation of polynomial expansions is computationally inexpensive, generating on the order of $10^5$ (or more) PCE-based samples is readily feasible.

\section*{Acknowledgment}

\par All the PCE models are constructed with the help of pygpc \cite{Weise2020Pygpc}. The sparse coefficients are determined using the LARS-Lasso solver from scipy \cite{Virtanen2020Scipy}.

\begin{IEEEbiography}[{\includegraphics[width=1in,height=1.25in,clip,keepaspectratio]{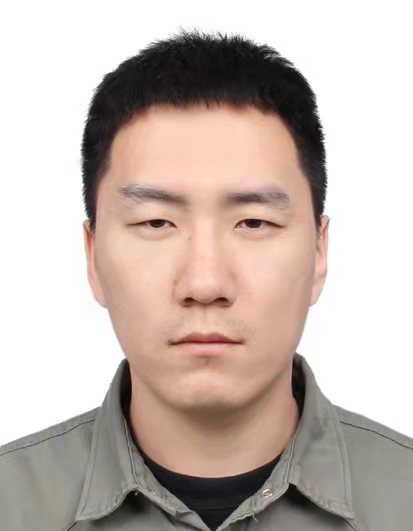}}]{Sicheng An} (Graduate Student Member, IEEE) received the B.E. degree in electronics and information engineering from Tiangong University, Tianjin, China, in 2020, and the M.S. degree in communication and information engineering from Harbin Institute of Technology, Harbin, China, in 2023. 
He is currently pursuing the Ph.D. degree in electrical engineering at Politecnico Di Milano, Milan, Italy.
He was a Research Assistant with the Department of Electrical and Computer Engineering, University of Alberta, Edmonton, Canada, from 2023 to 2024.
From 2024 to 2025, he was a Guest Researcher with the Max Planck Institute for Human Cognitive and Brain Sciences and Leipzig University of Applied Sciences (HTWK Leipzig), Leipzig, Germany.
His research interests include
uncertainty quantification and machine learning in computational electromagnetics, 
and individualized modeling of noninvasive brain stimulation.
He received the 2026 URSI Young Scientist Award.
\end{IEEEbiography}

\begin{IEEEbiography}[{\includegraphics[width=1in,height=1.25in,clip,keepaspectratio]{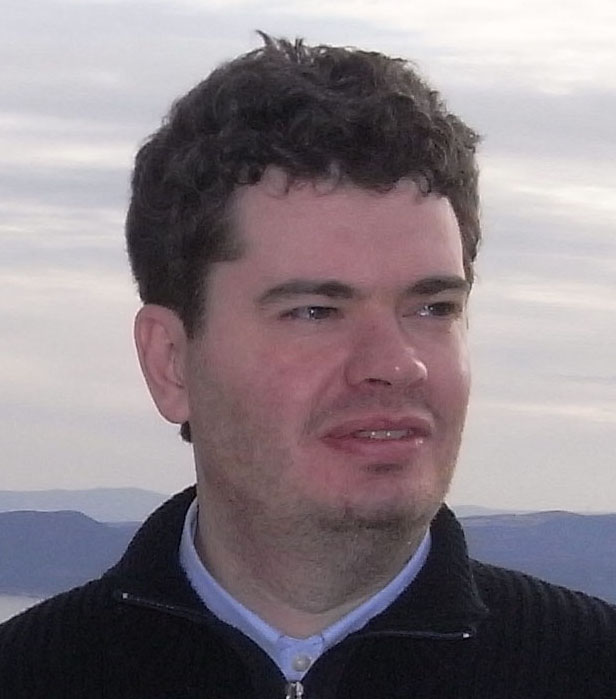}}]{Lorenzo Codecasa}(Member, IEEE) received the Ph.D. degree in electronics engineering from Politecnico di Milano, Milan, Italy, in 2001. 
Since 2002, he has been working first as a Researcher, then as an Associate Professor, and lastly as a 
Professor of electrical engineering with the DEIB Department of the same University.
In his research areas, he has authored or coauthored about 300 articles in refereed international
journals and conference proceedings. 
His main research contributions are in the theoretical analysis and the computational investigation of electronic circuits and electromagnetic fields.
In Stanford University's ranking, he is listed as one of the world's 2\% most influential scientists.
He is particularly active in the research of heat transfer and thermal management of electronic components and packages, in which he has been introducing original approaches to the extraction of reduced-order thermal models. 
Some of these techniques have become available in market-leading commercial software.
Dr. Codecasa received the Harvey Rosten Award for Excellence (twice) in 2015 and 2022 and four best papers awards at THERMINIC in 2014, 2017, 2019 and 2024. He has been an Associate Editor for IEEE TRANSACTIONS ON COMPONENTS, PACKAGING AND MANUFACTURING TECHNOLOGY since 2018. He has served as a Program Chair, the Vice-General Chair, and the General Chair of THERMINIC.
\end{IEEEbiography}

\begin{IEEEbiography}[{\includegraphics[width=1in,height=1.25in,clip,keepaspectratio]{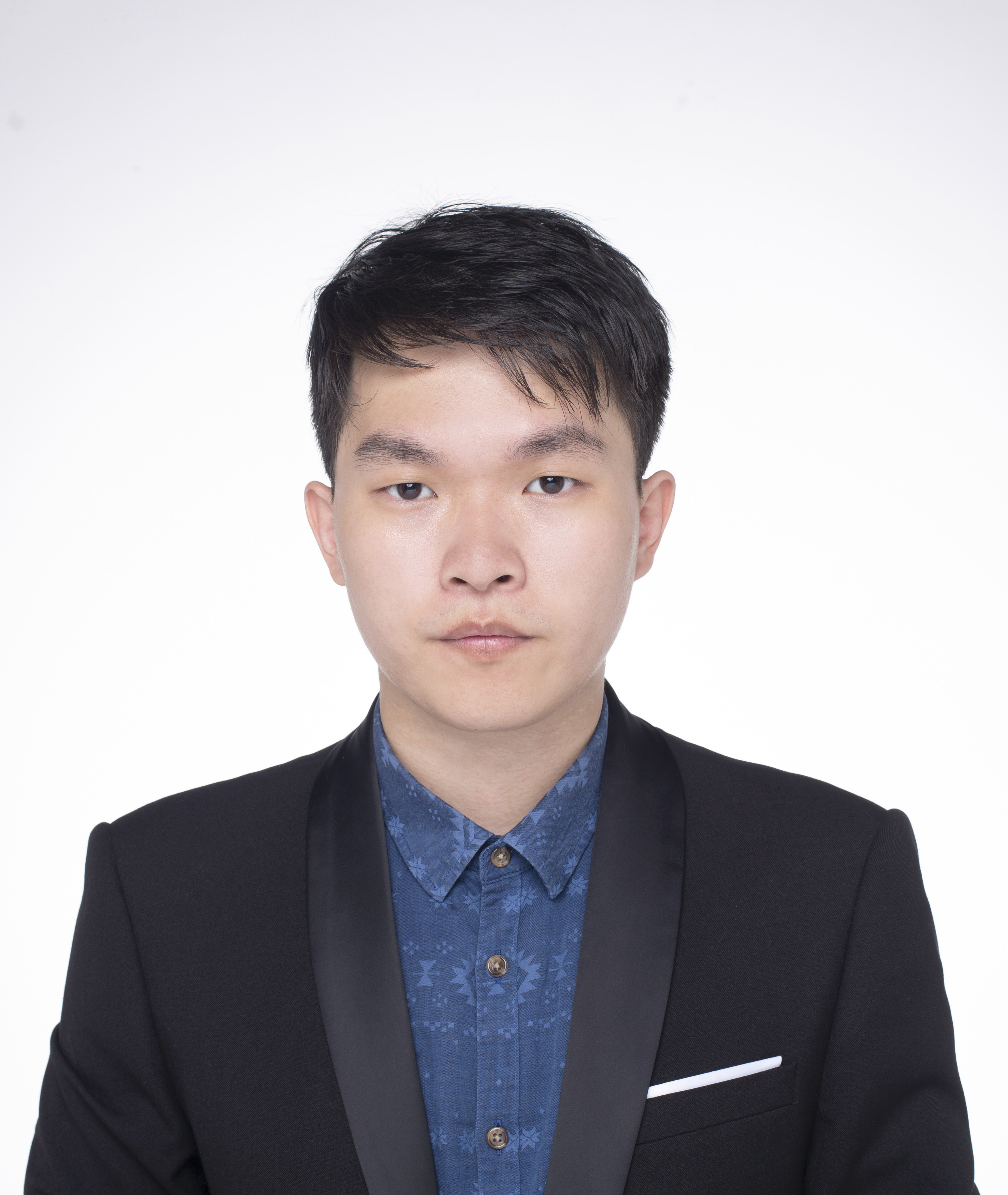}}]{Hao Qin} (Member, IEEE) received the B.E. and M.S. degrees in communication and information engineering from the Harbin Institute of Technology, Harbin, China, and the Ph.D. degree in electrical and electronic engineering from University College Dublin, Dublin, Ireland. He is currently an Assistant Professor at Sichuan University, Chengdu, China. His research interests include physics-informed machine learning for radio-wave propagation, computational electromagnetics, wireless channel modeling, and AI-native 6G systems. Dr. Qin was a recipient of the 2025 URSI Young Scientist Award, the 2023 Chinese Government Award for Outstanding Students Abroad, and the 2023 IEEE Antennas and Propagation Society Fellowship Program Award.
\end{IEEEbiography}

\begin{IEEEbiography}[{\includegraphics[width=1in,height=1.25in,clip,keepaspectratio]{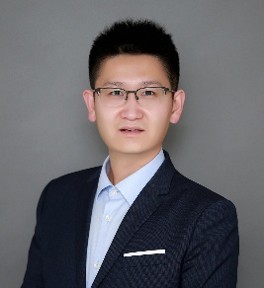}}]{Xingqi Zhang} (Senior Member, IEEE) received the B.Sc. degree from the Harbin Institute of Technology, China, in 2012, and the M.A.Sc. and Ph.D. degrees from the University of Toronto, Canada, in 2014 and 2018, respectively. He is currently an Associate Professor in the Department of Electrical and Computer Engineering, University of Alberta, Canada. He is also affiliated with the School of Electrical and Electronic Engineering, University College Dublin, Ireland, and has been a Visiting Professor at the University of Toronto and the Queen Mary University of London. His research is in the interdisciplinary areas of applied electromagnetics and wireless communications, with particular emphasis on computational electromagnetics, antennas and RF systems, radio wave propagation and wireless channel modeling, and AI-enabled electromagnetic and wireless system design. His research has applications in 5G/6G/THz wireless communications, intelligent transportation (air, ground, underground), industrial Internet of Things, as well as biomedical sensing and healthcare technologies.

He has served as a Distinguished Lecturer of the IEEE Vehicular Technology Society (VTS), and an Associate Editor for the IEEE Antennas and Wireless Propagation Letters (AWPL), the IET Microwaves, Antennas \& Propagation (MAP), and the IEEE Journal on Multiscale and Multiphysics Computational Techniques (JMMCT). In addition, he has been a Technical Program Committee Member and Session Chair for several flagship international conferences (e.g., IEEE AP-S/URSI, IEEE IWS, IEEE NEMO, IEEE VTC). His contributions have been recognized through numerous honors and awards, including the IEEE Sensors Council Technical Achievement Award in Sensor Systems or Networks, the Petro-Canada Emerging Innovator Award, the Applied Computational Electromagnetics Society (ACES) Early Career Award, the Royal Irish Academy (RIA) Charlemont Award, the Irish Research Council (IRC) Research Ally Prize, the IEEE ICCT Outstanding Young Scholar Award, five Young Scientist Awards from the International Union of Radio Science (URSI), ACES, and the Electromagnetics Academy, as well as several Best Paper Awards at international conferences and symposia.
\end{IEEEbiography}

\begin{IEEEbiography}[{\includegraphics[width=1in,height=1.25in,clip,keepaspectratio]{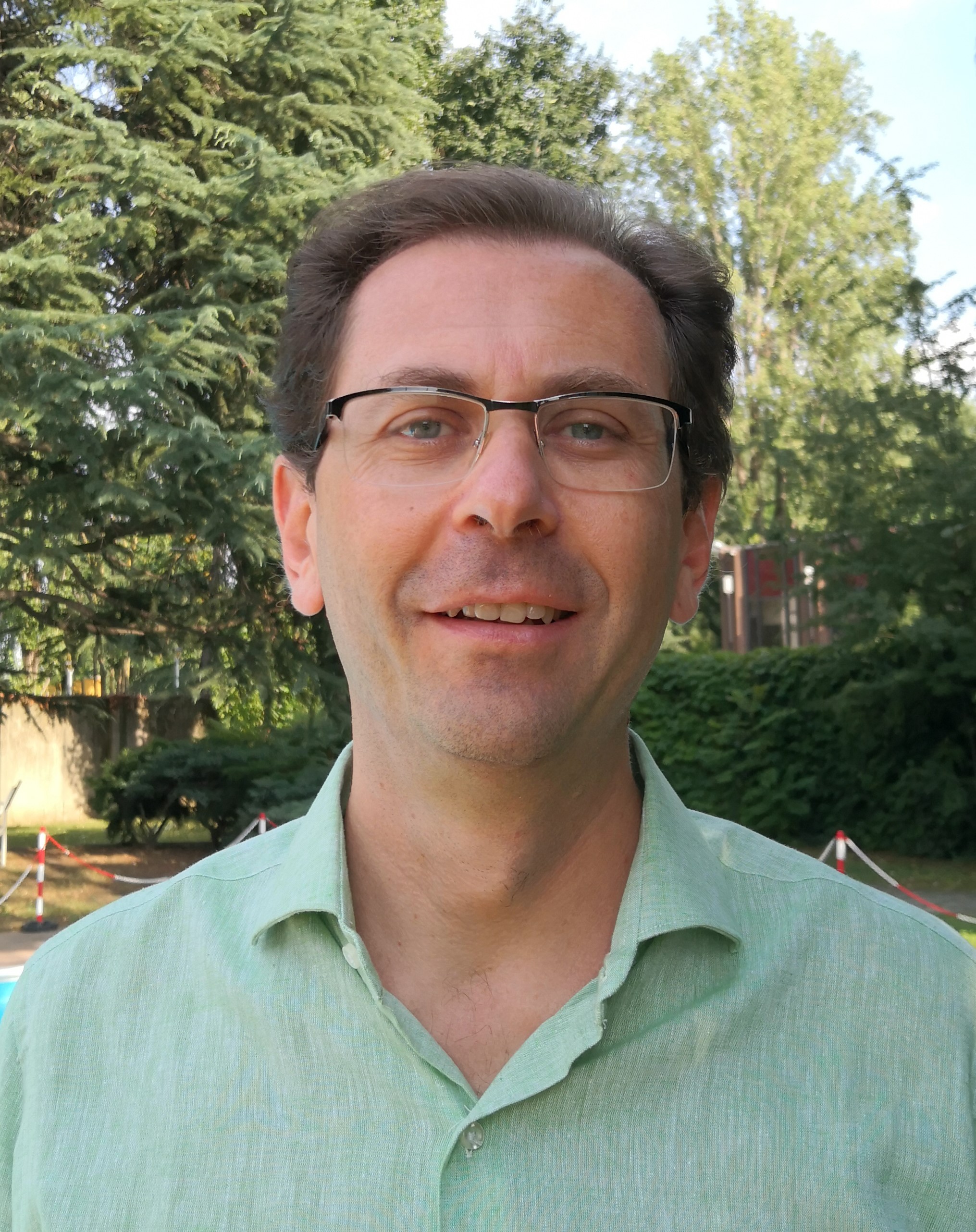}}]{Luca Di Rienzo}(Senior Member, IEEE) received the Laurea (M.Sc.) (cum laude) and Ph.D. degrees in electrical engineering from Politecnico di Milano (Milano, Italy) in 1996 and 2001, respectively, and the B.S. (cum laude) degree in mathematics from Università Statale di Milano (Milano, Italy) in 2020.
He is an Associate Professor with the Dipartimento di Elettronica, Informazione e Bioingegneria at Politecnico di Milano. His current research interests are in computational electromagnetics, including integral equation methods, surface impedance boundary conditions, model order reduction, and uncertainty quantification.
Prof. Di Rienzo is an Associate Editor of The Applied Computational Electromagnetics Society Journal, of the IEEE Open Access Journal of Power and Energy, and of IEEE Transactions on Magnetics, and a member of the Editorial Boards of COMPEL-The International Journal for Computation and Mathematics in Electrical and Electronic Engineering.
He is a co-author of more than 80 papers in international journals.

\end{IEEEbiography}


\begin{thebibliography}{00}

\bibitem{Yagbasan2010Characteristic}
A.~Yagbasan, C.~A.~Tunc, V.~B.~Erturk, A.~Altinatas, and R.~Mittra, 
``Characteristic basis function method for solving electromagnetic scattering problems over rough terrain profiles,'' 
\emph{IEEE Trans. Antennas Propag.}, 
vol.~58, no.~5, pp.~1579--1589, May~2010.

\bibitem{Huang2025AGeneralizable}
S.~Huang, H.~Qin, W.~Hou, X.~Zhang, and X.~Zhang, 
``A generalizable physics-guided convolutional neural network for irregular terrain propagation,'' 
\emph{IEEE Trans. Antennas Propag.}, 
vol.~73, no.~6, pp.~3975--3985, Jun.~2025.

\bibitem{Guo2017Pade}
Q.~Guo and Y. L. Long, ``Pade second-order parabolic equation modeling for propagation over irregular terrain,''
\emph{IEEE Antennas Wireless Propag. Lett.},
vol.~16, pp.~2853-2855, 2017

\bibitem{Batista2013Ahigh}
C.~G.~Batista and C.~G.~do Rego, 
``A high-order unconditionally stable FDTD-based propagation method,'' 
\emph{IEEE Antennas Wireless Propag. Lett.}, 
vol.~12, pp.~809--812, 2013.

\bibitem{Nuallain2005Terrain}
E.~O.~Nuallain,
``An efficient integral equation-based electromagnetic propagation model for terrain,''
\emph{IEEE Trans. Antennas Propag.}, vol.~53, no.~5, pp.~1836--1841, May 2005.

% \bibitem{Ozgun2009Recursive}
% O.~Ozgun, 
% ``Recursive two-way parabolic equation approach for modeling terrain effects in tropospheric propagation,'' 
% \emph{IEEE Trans. Antennas Propag.}, 
% vol.~57, no.~9, pp.~2706--2714, Sept.~2009.

\bibitem{Silva2012Terrain}
M.~A.~N.~Silva, E.~Costa, and M.~Liniger,
``Analysis of the effects of irregular terrain on radio wave propagation based on a three-dimensional parabolic equation,''
\emph{IEEE Trans. Antennas Propag.}, vol.~60, no.~4, pp.~2138--2143, 2012.

\bibitem{Qin2024Efficient}
H.~Qin, S.~An, and X.~Zhang, 
``Efficient two-way parabolic equation method with sparse Fourier transform for radio wave propagation over irregular terrain,'' 
in \emph{Proceedings of the Photonics \& Electromagnetics Research Symposium (PIERS)}, 
Chengdu, China, 2024, pp.~1--7.

\bibitem{Guo2020TWo}
Q.~Guo and Y.~Long, 
``Two-way parabolic equation method for radio propagation over rough sea surface,'' 
\emph{IEEE Trans. Antennas Propag.}, 
vol.~68, no.~6, pp.~4839--4847, Jun.~2020.

\bibitem{Qin2023Efficient}
H.~Qin and X.~Zhang, 
``Efficient radio wave propagation modeling in tunnels with a sparse Fourier transform-based split-step parabolic equation method,'' 
\emph{IEEE Antennas Wireless Propag. Lett.}, 
vol.~22, no.~10, pp.~2442--2446, Oct.~2023.

\bibitem{IEEEstd1492021}
IEEE Standards Association, 
\emph{IEEE Recommended Practice for Antenna Measurements}, 
IEEE Standard 149-2021, pp.~1--207, Feb.~2022.

\bibitem{ITU_R_F1336}
International Telecommunication Union, Radiocommunication Sector (ITU-R),  
``Reference radiation patterns of omnidirectional, sectoral and other antennas for the fixed and mobile services for use in sharing studies in the frequency range from 400~MHz to about 70~GHz,''  
Recommendation ITU-R~F.1336-5, Geneva, Switzerland, Jan.~2019.

\bibitem{Zhu2023Enhanced}
X. Zhu, L. Di Rienzo, X. Ma, and L. Codecasa, 
``Enhanced Multilevel Monte Carlo Method Applied to FDTD for Probability Distribution Estimation,''
\emph{IEEE Trans. Antennas Propag.}, vol. 71, no.10, pp.8390-8395, Oct.~2023.

\bibitem{An2024Efficient}
S.~An, H.~Qin, and X.~Zhang,  
``Efficient uncertainty quantification with subspace pursuit for FDTD-based microwave circuit models,''  
in \emph{Proceedings of the Photonics \& Electromagnetics Research Symposium (PIERS)},  
Chengdu, China, 2024, pp.~1--5.

% \bibitem{Austin2013Efficient}
% A.~C.~M.~Austin and C.~D.~Sarris,  
% ``Efficient analysis of geometrical uncertainty in the FDTD method using polynomial chaos with application to microwave circuits,''  
% \emph{IEEE Trans. Microwave Theory Tech.},  
% vol.~61, no.~12, pp.~4293--4301, 2013.

% \bibitem{Weise2015Uncertainty}
% K.~Weise, L.~Di~Rienzo, H.~Brauer, J.~Haueisen, and H.~Toepfer,  
% ``Uncertainty analysis in transcranial magnetic stimulation using nonintrusive polynomial chaos expansion,''  
% \emph{IEEE Trans. Magn.}, 
% vol.~51, no.~7, pp.~1--8, Jul.~2015.

\bibitem{Maitre2012Spectral}
O.~L.~Maitre and O.~M.~Knio,  
\emph{Spectral Methods for Uncertainty Quantification: With Applications to Computational Fluid Dynamics},  
Springer Netherlands, 2012.

\bibitem{Xiu2010Numerical}
D.~Xiu,  
\emph{Numerical Methods for Stochastic Computations: A Spectral Method Approach},  
Princeton, NJ, USA: Princeton University Press, 2010.

\rev{
\bibitem{ElMocayd2021Data}
N.~{El Mo\c{c}ayd} and M.~Seaid, 
``Data-driven polynomial chaos expansions for characterization of complex fluid rheology: Case study of phosphate slurry,'' 
\emph{Reliab. Eng. Syst. Saf.},
vol.~216, p.~107923, 2021.
}

\rev{
\bibitem{Codecasa2016Fast}
L.~Codecasa, L.~Di~Rienzo, K.~Weise, S.~Gross, and J.~Haueisen,  
``Fast MOR-based approach to uncertainty quantification in transcranial magnetic stimulation,''  
\emph{IEEE Trans. Magn.},  
vol.~52, no.~3, pp.~1--4, Mar.~2016.
}

\rev{
\bibitem{ElMocayd2020Stochastic}
N.~{El Mo\c{c}ayd}, M.~S.~Mohamed, D.~Ouazar, and M.~Seaid,
``Stochastic model reduction for polynomial chaos expansion of acoustic waves using proper orthogonal decomposition,''
\emph{Reliab. Eng. Syst. Saf.}, 
vol.~195, p.~106733, 2020.
}

\bibitem{Luthen2021sparse}
N.~Lüthen, S.~Marelli, and B.~Sudret,  
``Sparse polynomial chaos expansions: Literature survey and benchmark,''  
\emph{SIAM/ASA J. Uncertain. Quantif.},  
vol.~9, no.~2, pp.~593--649, 2021.

\bibitem{Blatman2011Adaptive}
G.~Blatman and B.~Sudret,  
``Adaptive sparse polynomial chaos expansion based on least angle regression,''  
\emph{J. Comput. Phys.},  
vol.~230, no.~6, pp.~2345--2367, 2011.

\bibitem{Hampton2018Basis}
J.~Hampton and A.~Doostan, 
``Basis adaptive sample efficient polynomial chaos ({BASE-PC}),'' 
\emph{J. Comput. Phys.}, 
vol.~371, pp.~20--49, 2018.

\bibitem{Luthen2022Automatic}
N.~Lüthen, S.~Marelli, and B.~Sudret,  
``Automatic selection of basis-adaptive sparse polynomial chaos expansions for engineering applications,''  
\emph{Int. J. Uncertain. Quantif.},  
vol.~12, no.~3, 2022.

\bibitem{Hao2025Combined}
D.~Hao, J.~Zhang, X.~Yue, and L.~Chen,  
``Combined dimensionality reduction based adaptive polynomial chaos expansion for high-dimensional reliability analysis,''  
\emph{Reliab. Eng. Syst. Saf.},  
no.~111324, 2025.

\bibitem{Jakeman2015Enhancing}
J.~D.~Jakeman, M.~S.~Eldred, and K.~Sargsyan,
``Enhancing l\textsubscript{1}-minimization estimates of polynomial chaos expansions using basis selection,''
\emph{J. Comput. Phys.}, vol.~289, pp.~18--34, 2015.

\bibitem{Saturnino2019A}
G.~B.~Saturnino, A.~Thielscher, K.~H.~Madsen, T.~R.~Knösche, and K.~Weise,  
``A principled approach to conductivity uncertainty analysis in electric field calculations,''  
\emph{Neuroimage},  
vol.~188, pp.~821--834, 2019.

\bibitem{An2025Multilevel}
S.~An, L.~Di~Rienzo, H.~Qin, X.~Zhu, X.~Zhang, and L.~Codecasa,
``Multilevel Monte Carlo coupled with the parabolic wave equation method for uncertainty analysis of radio wave propagation in tunnels,'' 
\emph{IEEE Trans. Antennas Propag.},
vol.~74, no.~2, pp.~2191--2196, Feb. 2026.

\bibitem{Zhang2019Statistical}
X.~Zhang, N.~Sood, J.~K.~Siu, and C.~D.~Sarris,  
``Statistical modeling of electromagnetic wave propagation in tunnels with rough walls using the vector parabolic equation method,''  
\emph{IEEE Trans. Antennas Propag.},  
vol.~67, no.~4, pp.~2645--2654, 2019.

% \bibitem{An2026Probability}
% S.~An, Z.~Ye, L.~Di Rienzo, H.~Qin, X.~Zhang and X.~Zhang, 
% ``Probability distribution function estimation for tunnel propagation via the multilevel Monte Carlo method,'' 
% in \emph{Proc. IEEE Int. Conf. Comput. Electromagn. (ICCEM)},
% Shanghai, China, 2026, pp.~1--4.

% \bibitem{Gerstner2003Dimension}
% T.~Gerstner and M.~Griebel,  
% ``Dimension-adaptive tensor-product quadrature,''  
% \emph{Computing},  
% vol.~71, no.~1, pp.~65--87, 2003.


% \bibitem{Papadopoulos2021Anisotropic}
% A.~Papadopoulos, T.~Zygiridis, E.~Glytsis, and N.~Kantartzis,
% ``An adaptive sparse anisotropic polynomial-chaos expansion algorithm applied to EMC problems,''
% \emph{IEEE Electromagn. Compat. Mag.}, vol.~10, no.~2, pp.~80--87, 2021.


% \bibitem{Zhang2023Anovel}
% B.~Y.~Zhang and Y.~Q.~Ni,
% ``A novel sparse polynomial chaos expansion technique with high adaptiveness for surrogate modelling,''
% \emph{Appl. Math. Model.}, vol.~121, pp.~562--585, 2023.


% \bibitem{Balanis1989Advanced}
% C.~A.~Balanis, \emph{Advanced Engineering Electromagnetics},
% Hoboken, NJ, USA: Wiley, 1989.

\bibitem{Levy2000Parabolic}
M.~Levy,  
\emph{Parabolic Equation Methods for Electromagnetic Wave Propagation},  
London, U.K.: IEE, 2000.

% \bibitem{Tropp2007Signal}
% J.~A.~Tropp and A.~C.~Gilbert,  
% ``Signal recovery from random measurements via orthogonal matching pursuit,''  
% \emph{IEEE Transactions on Information Theory},  
% vol.~53, no.~12, pp.~4655--4666, 2007.

% \bibitem{Sobol2001Global}
% I.~M.~Sobol,  
% ``Global sensitivity indices for nonlinear mathematical models and their Monte Carlo estimates,''  
% \emph{Math. Comput. Simul.},  
% vol.~55, no.~1–3, pp.~271--280, 2001.

% \bibitem{Sudret2008Global}
% B.~Sudret,  
% ``Global sensitivity analysis using polynomial chaos expansions,''  
% \emph{Reliab. Eng. Syst. Saf.},  
% vol.~93, no.~7, pp.~964--979, 2008.

\bibitem{Blatman2008Sparse}
G.~Blatman and B.~Sudret,  
``Sparse polynomial chaos expansions and adaptive stochastic finite elements using a regression approach,''  
\emph{C. R. Méc.},  
vol.~336, no.~6, pp.~518--523, 2008.

\bibitem{Xiu2002TheWiener}
D.~Xiu and G.~E.~Karniadakis,  
``The Wiener–Askey polynomial chaos for stochastic differential equations,''  
\emph{SIAM J. Sci. Comput.},  
vol.~24, no.~2, pp.~619--644, 2002.

% \bibitem{Helton2003Latin}
% J.~C.~Helton and F.~J.~Davis,  
% ``Latin hypercube sampling and the propagation of uncertainty in analyses of complex systems,''  
% \emph{Reliab. Eng. Syst. Saf.},  
% vol.~81, no.~1, pp.~23--69, 2003.

% \bibitem{Weise2022Comparison}
% K.~Weise, E.~Müller, L.~Poßner, and T.~R.~Knösche,  
% ``Comparison of the performance and reliability between improved sampling strategies for polynomial chaos expansion,''  
% \emph{Mathematical Biosciences and Engineering},  
% vol.~19, no.~8, pp.~7425--7480, 2022.

\bibitem{BlatmanThesis}
G.~Blatman,
``Adaptive sparse polynomial chaos expansions for uncertainty propagation and sensitivity analysis,'' 
Ph.D. dissertation, Universit\'{e} Blaise Pascal, Clermont-Ferrand, France, 2009.

\bibitem{Loukrezis2025Multivariate}
D.~Loukrezis, E.~Diehl, and H.~De~Gersem,
``Multivariate sensitivity-adaptive polynomial chaos expansion for high-dimensional surrogate modeling and uncertainty quantification,''
\emph{Appl. Math. Model.}, vol.~137, 2025, Art. no. 115746.

% \bibitem{Hampton2018Basis}
% J.~Hampton and A.~Doostan,
% ``Basis adaptive sample efficient polynomial chaos (BASE-PC),''
% \emph{J. Comput. Phys.}, vol.~371, pp.~20--49, 2018.

\bibitem{Wegert2025Conductivity}
L.~Wegert, L.~Di~Rienzo, L.~Codecasa, S.~An, M.~Ziolkowski, A.~Hunold, I.~Lange, T.~Kalla, and J.~Haueisen,
``The influence of tissue conductivity uncertainty on the nerve activation thresholds in non-invasive electrical phrenic nerve stimulation,''
\emph{Biocybern. Biomed. Eng.}, vol.~45, no.~4, pp.~697--706, 2025.

\bibitem{Hviid1995Terrain}
J.~T.~Hviid, J.~B.~Andersen, J.~Toftgard, and J.~Bojer,  
``Terrain-based propagation model for rural area—An integral equation approach,''  
\emph{IEEE Trans. Antennas Propag.},  
vol.~43, no.~1, pp.~41--46, Jan.~1995.

\bibitem{Qin2024Comparative}
H.~Qin and X.~Zhang,  
``Comparative analysis of finite-difference and split-step based parabolic equation methods for tunnel propagation modelling,''  
\emph{IET Microw. Antennas Propag.},  
vol.~18, no.~2, pp.~59--72, 2024.

\bibitem{Gupta2004Handbook}
A.~K.~Gupta and S.~Nadarajah,  
\emph{Handbook of Beta Distribution and Its Applications},  
New York, NY, USA: Marcel Dekker, 2004.

\bibitem{Apaydin2017Radio}
G.~Apaydin and L.~Sevgi, \emph{Radio Wave Propagation and Parabolic Equation Modeling}. John Wiley \& Sons, 2017.

\bibitem{Weise2020Pygpc}
K.~Weise, L.~Poßner, E.~Müller, R.~Gast, and T.~R.~Knösche,  
``Pygpc: A sensitivity and uncertainty analysis toolbox for Python,''  
\emph{SoftwareX}, vol.~11, 2020.

\bibitem{Virtanen2020Scipy}
P.~Virtanen, R.~Gommers, T.~E.~Oliphant, M.~Haberland, T.~Reddy, D.~Cournapeau, \emph{et al.},  
``SciPy 1.0: Fundamental algorithms for scientific computing in Python,''  
\emph{Nat. Methods},  
vol.~17, pp.~261--272, 2020.



\end{thebibliography}
\end{document}